\documentclass[twocolumn,superscriptaddress,amsmath,amssymb,floatfix,aps,notitlepage,prx,longbibliography]{revtex4-1}
\usepackage{mathrsfs, hyperref}
\usepackage{amssymb, amsbsy, amsmath, latexsym, dsfont, array, layout, graphics,mathrsfs,braket,amsfonts,amsthm,
  amssymb,graphicx,subfigure, youngtab,color,bm,mathtools,braket,verbatim,url,cleveref,natbib,hypernat,extarrows,inputenc,multirow}
\usepackage[usernames,dvipsnames]{xcolor}

\begin{document}
\title{Buckling of thermalized elastic sheets}

 \author{Ali Morshedifard}
 \affiliation{Department of Civil and Environmental Engineering, Henry Samueli School of Engineering, E4130 Engineering Gateway, University of California, Irvine, Irvine, CA 92697-2175 USA.}

 \author{Miguel Ruiz-Garc\'{i}a}
 \affiliation{Department of Physics and Astronomy, University of Pennsylvania, Philadelphia, PA 19104, USA}
 
  \author{Mohammad Javad Abdolhosseini Qomi}
 \email{mjaq@uci.edu}
 \affiliation{Department of Civil and Environmental Engineering, Henry Samueli School of Engineering, E4130 Engineering Gateway, University of California, Irvine, Irvine, CA 92697-2175 USA.}

 \author{Andrej Ko\v{s}mrlj}
 \email{andrej@princeton.edu}
 \affiliation{Department of Mechanical and Aerospace Engineering, Princeton University, Princeton, NJ 08544, USA}
 \affiliation{Princeton Institute for the Science and Technology of Materials, Princeton University, Princeton, NJ 08544, USA}


\begin{abstract}
Steady progress in the miniaturization of structures and devices has reached a scale where thermal fluctuations become relevant and it is thus important to understand how such fluctuations affect their mechanical stability. Here, we investigate the buckling of thermalized sheets and we demonstrate that thermal fluctuations increase the critical buckling load due to the enhanced scale-dependent bending rigidity for sheets that are much larger than a characteristic thermal length scale. The presented results are universal and apply to a wide range of microscopic sheets. These results are especially relevant for atomically thin 2D materials, where thermal fluctuations can significantly increase the critical buckling load because the thermal length scale is on the order of nanometers at room temperature.
\end{abstract}

\maketitle

The mechanics of slender structures has been studied for centuries~\cite{timoshenkoBhistory} but is still actively explored to this day because geometrical nonlinearities lead to many interesting phenomena involving buckling, wrinkling, and folding~\cite{brau13,chopin13,stoop15,paulsen16,nagashima17}. In the 19th century, a concentrated effort was made to characterize critical loads at the onset of mechanical instabilities~\cite{timoshenkoBShells,koiterB,novozhilovB,ciarletB,landauB}, because engineers had to design stable and safe buildings, structures, and machines. However, in recent years, it has become a trend to exploit these instabilities in order to make so-called mechanical metamaterials in a wide range of applications including flexible electronics~\cite{wongB,shyu15,xu15}, flexible photovoltaics~\cite{pagliaroB,schubert06,docampo13,lamoureux15}, tunable surface properties (drag, adhesion, hydrophobicity/hydrophilicity)~\cite{lin08,yang10,terwagne14}, tunable photonic and phononic band gaps~\cite{kim01,wong04,li12,wang13}, mechanical cloaks~\cite{buckmann14,buckmann15}, self-assembled/self-folded robots and structures~\cite{pandey11,cho11,felton14}, shape-changing materials~\cite{klein07,kim12Science,gladman16}, and mechanical topological metamaterials~\cite{sun12,kane13,paulose15,nash15,rocklin16}. 

Many slender structures and mechanical metamaterials have been successfully translated from the macroscopic to the microscopic scale, where they are used as flexible electronics~\cite{blees15, zhang15}, self-folding structures~\cite{leong09,malachowski14,xu2017ultrathin, miskin2018graphene, reynolds2019capillary}, devices for targeted drug delivery~\cite{li16,kagan10,gao12,wu13AngewChem}, for the manipulation and isolation of cells~\cite{balasubramanian11,sanchez11,petit12}, and also in diverse environmental and industrial applications, including water monitoring, remediation, and detoxification~\cite{gao14,soler14,singh15}. As we strive to make devices and machines smaller and smaller, we are ultimately going to reach a scale where defects and thermal fluctuations become relevant. Thus, it is important to characterize how these two effects are going to affect mechanical properties and the stability of slender structures. In this letter, we focus on thermal fluctuations that become relevant once the narrow dimensions of structures reach the scale of nanometers. In many systems this condition is already satisfied, such as for graphene kirigami~\cite{blees15} and graphene-based self-folding origami~\cite{xu2017ultrathin, miskin2018graphene}, where we expect that thermal fluctuations significantly affect their mechanical properties, including the critical buckling load, which is the focus of this letter.

\begin{figure}[t!]
    \centering
    \includegraphics[width=0.5\textwidth]{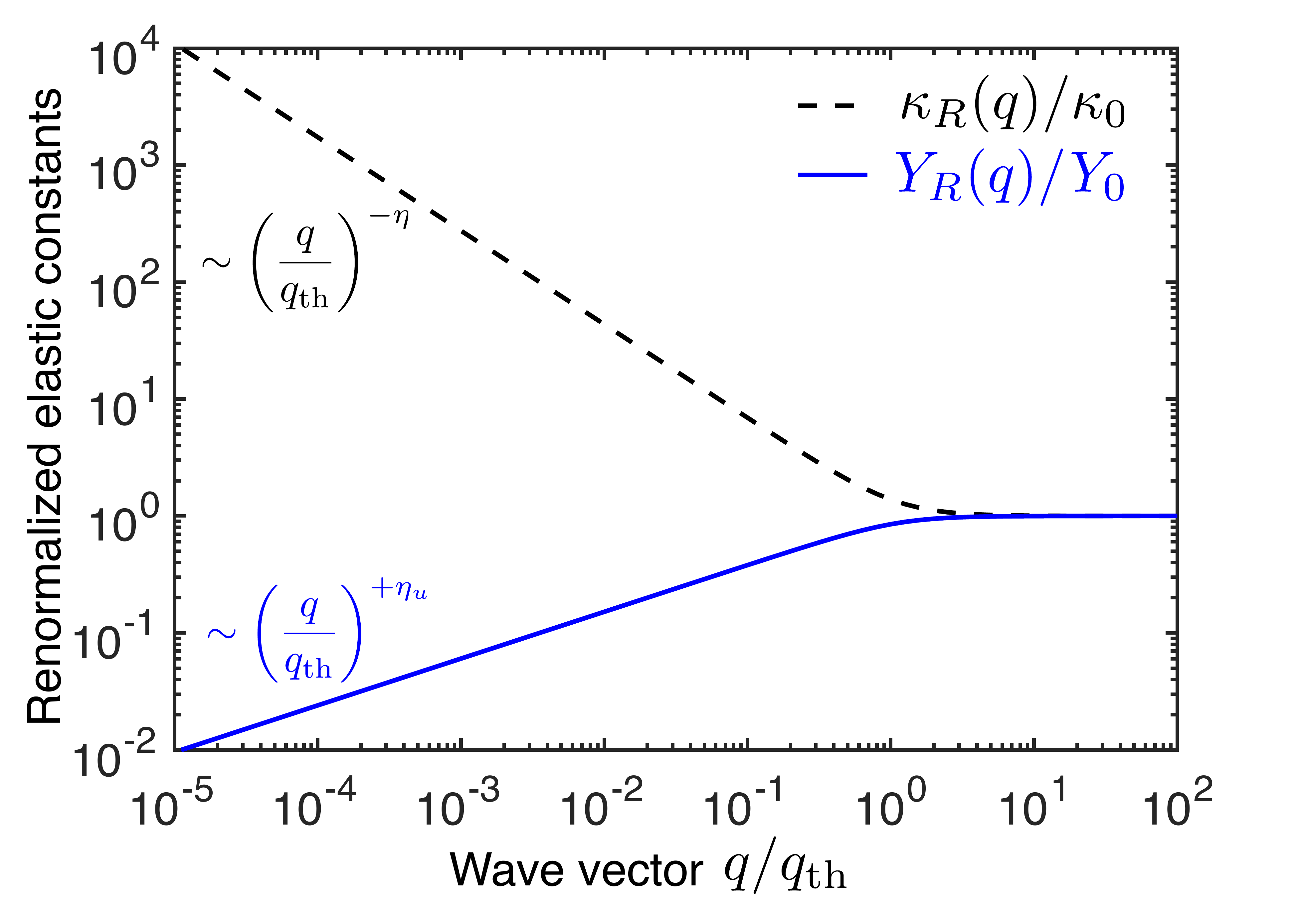}
    \caption{Universal scaling functions for the renormalized bending rigidity $\kappa_R(q)$ and Young's modulus $Y_R(q)$ that are normalized by the zero temperature values $\kappa_0$ and $Y_0$. The temperature dependent transition wave vector $q_\text{th}$ is defined in Eq.~(\ref{eq:qth}). Scaling functions were obtained with the perturbative renormalization group procedure~\cite{kosmrlj16}.}
    \label{fig1}
\end{figure}

Here, we consider a coarse-grained description of a freely suspended linear elastic sheet with the bending rigidity $\kappa_0$ and 
the 2D Young's modulus $Y_0$. In the absence of external loads, thermal fluctuations effectively modify elastic constants and make them scale dependent~\cite{nelsonB,katsnelsonB,amorim16}. We refer to these effective constants as the renormalized elastic constants. The renormalized bending rigidity $\kappa_R$ can be extracted from the spectrum of the height fluctuations $h({\bf r})$, which are the out-of-plane displacements from the reference undeformed athermal flat state and they can  be measured in experiments or simulations, as~\cite{nelsonB,katsnelsonB,amorim16}
\begin{eqnarray}
\langle h({\bf q})h(-{\bf q})\rangle &=& \frac{k_B T}{A \kappa_R(q) q^4},
\end{eqnarray}
where we introduced the Boltzmann constant $k_B$, the ambient temperature $T$, the area $A$ of the undeformed flat sheet,
and the Fourier transform of the height profile $h({\bf q}) = \int \! (d^2{\bf r}/A) \, e^{-i {\bf q} \cdot {\bf r}} h({\bf r})$. Here, ${\bf r}\equiv(x,y)$ and ${\bf q}\equiv(q_x,q_y)$.
Similarly, the renormalized Young's modulus $Y_R$ can be obtained from the fluctuation spectrum of the in-plane displacements~\cite{nelsonB,katsnelsonB,amorim16,kosmrlj16}. Figure~\ref{fig1} shows the scaling functions for the renormalized bending rigidity $\kappa_R(q)$ and Young's modulus $Y_R(q)$.  Thermal fluctuations effectively increase the bending rigidity and reduce the Young's modulus, which scale as
\begin{eqnarray}
\frac{\kappa_R(q)}{\kappa_0} &\sim&  \left\{
\begin{array}{c l}
1, & q \gg q_\textrm{th} \\
(q/ q_\textrm{th})^{-\eta}, & q 
\ll q_\textrm{th} \\
\end{array}
\right. , \\
\frac{Y_R(q)}{Y_0} &\sim& \left\{
\begin{array}{c l}
1, & q \gg q_\textrm{th} \\
(q/q_\textrm{th})^{+\eta_u}, & q \ll q_\textrm{th} \\
\end{array}
\right. . \nonumber
\label{eq:renormalized_elastic_constants}
\end{eqnarray}
Here the scaling exponents $\eta\approx 0.80-0.85$ and $\eta_u=2-2 \eta\approx 0.3-0.4$, which were estimated theoretically~\cite{nelson87,aronovitz88,guitter88,guitter89,aronovitz89,nelson91,ledoussal92} and confirmed in atomistic and coarse-grained Monte Carlo simulations~\cite{zhang93,bowick96,bowick97,los09,roldan11,troster13,troster15,los16}, quantify the scale dependence of the elastic constants driven by thermal fluctuations in the range of wave vectors up to the transition scale~\cite{aronovitz88,guitter88,guitter89,nelson91}
\begin{equation}
    q_\text{th}=\sqrt{\frac{3 k_B TY_0}{16 \pi \kappa_0^2}}.
    \label{eq:qth}
\end{equation}
above which thermal fluctuations are no longer significant. This transition scale can be used to define the thermal length scale
\begin{equation}
\ell_\text{th}\equiv \frac{2 \pi}{ q_\text{th}} = \sqrt{\frac{64 \pi^3 \kappa_0^2}{3 k_B TY_0}}
\label{eq:thermal_length}
\end{equation}
beyond which thermal fluctuations become important.

The scaling functions for the renormalized bending rigidity $\kappa_R(q)$ and Young's modulus $Y_R(q)$ presented in Fig.~\ref{fig1} are universal and they hold for all isotropic microscopic sheets, where the continuum theory applies, i.e. for wave vectors $q$ that are much smaller than the microscopic cutoff $\Lambda \sim 1/a_0$, where $a_0$ is e.g. the lattice spacing in 2D crystalline sheets. Note that due to thermal fluctuations the microscopically anisotropic sheets, such as black phosphorene, behave like isotropic sheets in the long wavelength limit~\cite{toner1989elastic}. At room temperatures thermal fluctuations are important for freely suspended 2D crystalline sheets, such as graphene ($\ell_\text{th}\approx 4\,\text{nm}$~\cite{fasolino07,lee08}) or  MoS$_2$ ($\ell_\text{th}\approx 40-50\,\text{nm}$~\cite{bertolazzi2011stretching,lai16}), which can easily be fabricated on the microscale. Thus thermal fluctuations in these systems significantly increase the bending rigidity and reduce the Young's modulus (see Fig.~\ref{fig1}). The characteristic diameter of clay platelets is $\approx 100-500\,\text{nm}$, which is comparable to the thermal length scale $\ell_\text{th}\approx 100-1,000\,\text{nm}$ at room temperature~\cite{suter2007}. Similarly, the diameter of red blood cells  ($6-8\,\mu\text{m}$) is comparable to the thermal length scale  $\ell_\text{th}\approx 2-10\,\mu\text{m}$ at room temperature~\cite{waugh79, evans83, park10,evans17}.

When the external compressive load $\sigma_{ij}$ ($i,j\in\{x,y\}$) is applied to the boundary of the elastic sheet, the spectrum of height fluctuations becomes~\cite{roldan11,kosmrlj16}
\begin{eqnarray}
\langle h({\bf q})h(-{\bf q})\rangle &=& \frac{k_B T}{A \left[\kappa_R(q) q^4-\sigma_{ij} q_i q_j\right]}.
\label{eq:height_fluctuations}
\end{eqnarray}
Note that for sufficiently large compressive loads $\sigma_{ij}$ the denominator in Eq.~(\ref{eq:height_fluctuations}) can become negative, which reflects the fact that the flat state becomes unstable and the sheet buckles. The critical buckling load $\sigma_R^b$ corresponds to the minimum compressive load, where the denominator in Eq.~(\ref{eq:height_fluctuations}) vanishes. For the biaxial compression ($\sigma_{ij}=\sigma \delta_{ij}$, where $\delta_{ij}$ is the Kronecker delta) and for the uniaxial compression ($\sigma_{ij}=\sigma \delta_{ix} \delta_{jx}$) the critical buckling load is 
$\sigma_R^b=\kappa_R(q_\text{min}) q_\text{min}^2$,
where $q_\text{min}$ is the smallest wave vector allowed by the boundary conditions. 

Here, we consider periodic boundary conditions for the biaxial and uniaxial compression, as well as the clamped-free boundary conditions for the uniaxial compression (two edges that experience the load are clamped, while the other two edges are free). For a square sheet of size $\ell_0 \times \ell_0$ the smallest allowed wave vector is $q_\text{min}=2 \pi /\ell_0$ for all considered cases. The critical buckling load $\sigma_R^b$ thus scales as 
\begin{eqnarray}
\sigma_R^b = \kappa_R(q_\text{min}) q_\text{min}^2  \sim \left\{
\begin{array}{c l}
\kappa_0 \ell_0^{-2}, & \ell_0 \ll \ell_\textrm{th} \\
\kappa_0 \ell_0^{-2+\eta} \ell_\textrm{th}^{-\eta}, & \ell_0
\gg \ell_\textrm{th} \\
\end{array}
\right. .
\label{eq:buckling_load}
\end{eqnarray}
Note that the critical buckling load is temperature dependent and scales as $\sigma_R^b\sim T^{\eta/2}$ for elastic sheets that are larger than the thermal length scale ($\ell_0 \gg \ell_\text{th}$), e.g. graphene, MoS$_2$, and other 2D crystalline materials.
 Compared to the classical value for the critical buckling load $\sigma_0^b = 4 \pi \kappa_0 \ell_0^{-2}$ at zero temperature~\cite{timoshenkoBShells,koiterB}, thermal fluctuations effectively increase the critical buckling load due to the enhanced renormalized bending rigidity as
\begin{equation}
\frac{\sigma_R^b}{\sigma_0^b} = \frac{\kappa_R(q_\text{min})}{\kappa_0} \equiv \frac{\overline{\kappa}_R(\ell_0)}{\kappa_0}.
\label{eq:normalized_buckling_load}
\end{equation}
Note that this is different from spherical shells, where thermal fluctuations effectively reduce the critical buckling pressure~\cite{paulose12, kosmrlj17, baumgarten2018buckling,singh2020finite}. Note also that the applied external load could affect the renormalization of the bending rigidity as was previously demonstrated for sheets under tension~\cite{guitter89,morse92,radzihovsky98,kosmrlj16,burmistrov2018stress}. This effect becomes important when the contribution from external load in the denominator of Eq.~(\ref{eq:height_fluctuations}) becomes dominant, which happens only after the sheet buckles. Thus we expect that the renormalization of the bending rigidity is not significantly affected up to the critical buckling load, but it is likely affected in the post-buckling regime.

\begin{figure}[t!]
    \centering
    \includegraphics[width=0.5\textwidth]{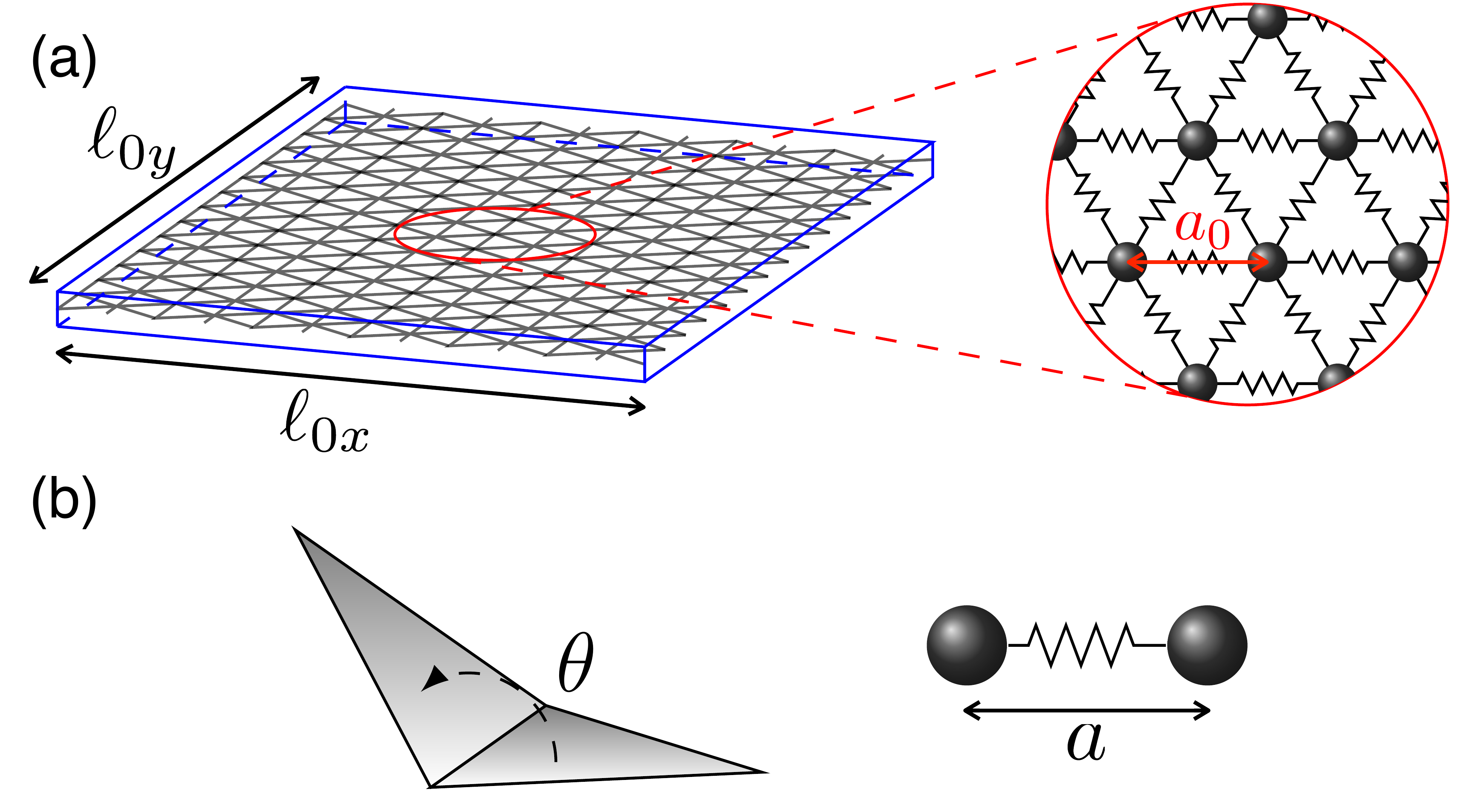}
    \caption{(a) An elastic sheet is represented as an equilateral triangulation of a rectangle with size $\ell_{0x} \times \ell_{0y}$  ($\ell_{0x} \approx \ell_{0y}$) where particles are positioned on lattice points 
    with bending
and stretching energies defined along the edges and plaquettes of
the triangulation. (b) The bending energy is described as a penalty of changing the dihedral angle $\theta$
between two triangles sharing an edge. The stretching energy is described as a penalty of changing the bond length $a$ between the two particles.}
    \label{fig2}
\end{figure}

To test the prediction for the critical buckling load in Eq.~(\ref{eq:height_fluctuations}) we performed coarse-grained Molecular Dynamics simulations, where the elastic sheet is discretized as a triangulation of a nearly square sheet of size $\ell_{0x} \times \ell_{0y}$ with $\ell_{0x} \approx \ell_{0y}$ (see Fig.~\ref{fig2}a). Such simulations were previously used to test the renormalization of elastic constants~\cite{zhang93,bowick96,bowick97,bowick17} and they agreed very well with both the continuum theory and the atomistic Monte Carlo simulations~\cite{los09,roldan11,los16} on scales that are much larger than the lattice constant and interatomic spacing.

In the undeformed flat configuration, all triangles are equilateral with the edge length $a_0$. The bending and stretching energies are described using a common discretization~\cite{seung88} of the continuum energy as
\begin{equation}
\begin{split}
{U_\text{bend}} &=  \sum_{<I,J>} k_\text{bend}{(1 + \cos \theta_{IJ} )},\\
{U_\text{stretch}} &=  {\sum_{<i,j>} \frac{1}{2}k_\text{stretch} {({a_{ij}} - a_0)} ^2},
\end{split}
\label{eq:energy} 
\end{equation}
where $\theta_{IJ}$ is the dihedral angle between the neighbor triangles $I$ and $J$ that are sharing an edge, and $a_{ij}=|{\bf r}_i-{\bf r}_j|$ is the Euclidean distance between the nearest-neighbor particles $i$ and $j$. Note that the discretization parameters $k_\text{bend}$ and $k_\text{stretch}$ are directly related to the continuum bending rigidity $\kappa_0 = \frac{\sqrt{3}}{2} k_\text{bend}$, the continuum Young's modulus $Y_0 = \frac{2}{\sqrt{3}} k_\text{stretch}$, and the continuum Poisson's ratio $\nu_0=1/3$~\cite{seung88,schmidt12}.

Molecular  Dynamics  simulations  were performed using the LAMMPS package~\cite{LAMMPS} (see Appendix for details). Unless otherwise noted the undeformed size of the sheet was $\ell_0= 100\,a_0$ ($\ell_{0x}=116\, a_0 \sqrt{3}/2 \approx 100.5\, a_0$ and  $\ell_{0y}=100\,a_0$). To adjust the thermal length scale $\ell_\text{th}\sim\kappa_0/\sqrt{k_B T Y_0}$ we varied the temperature $T$, and the bending and stretching spring constants $k_\text{bend}$ and $k_\text{stretch}$, respectively, which enabled us to explore a wide range of thermal length scales  ($\ell_0/\ell_\text{th}\in (10^{-2},10^5)$).

\begin{figure*}[t!]
\begin{centering}
\includegraphics[width=.8\textwidth]{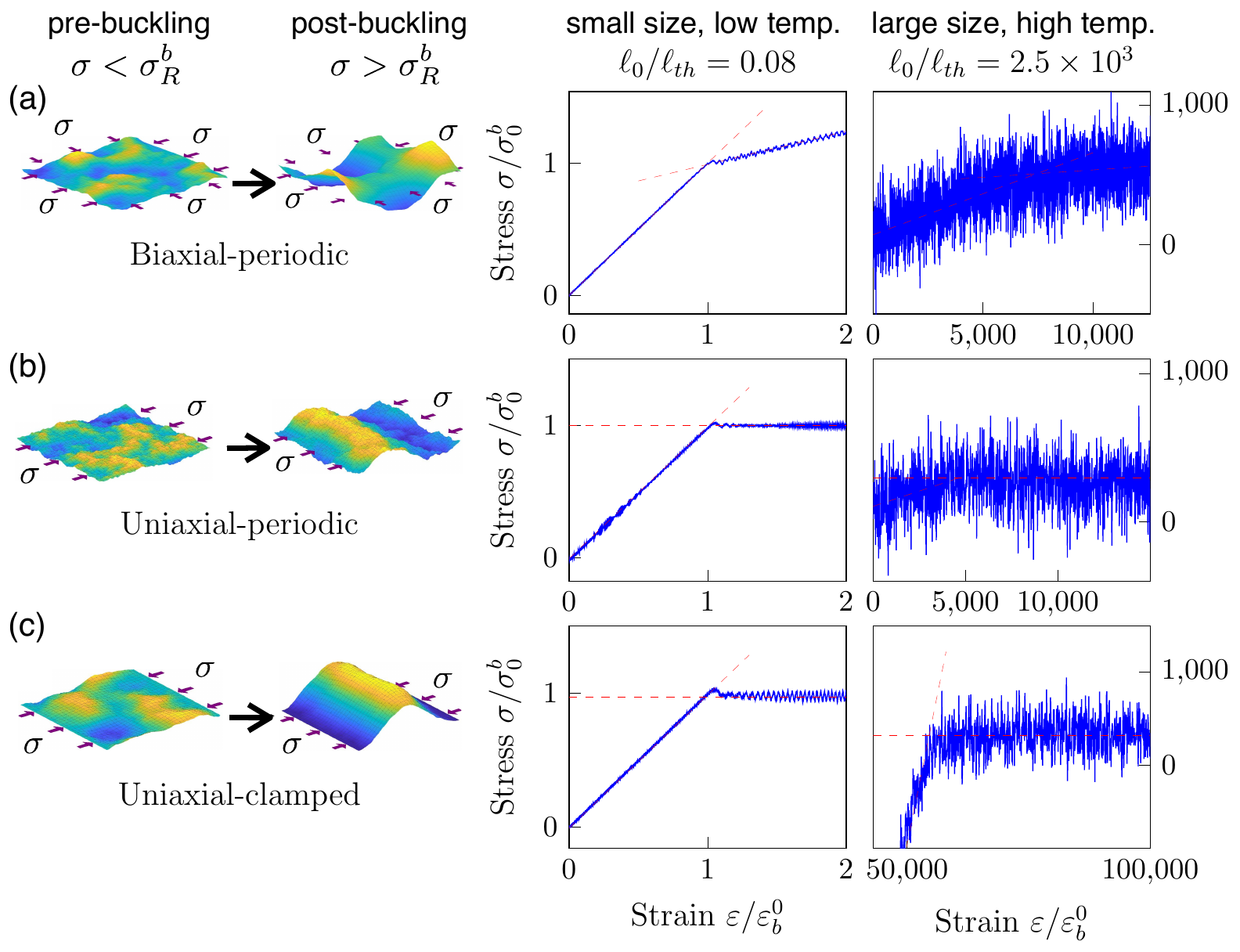}
\end{centering}
\caption{Representative simulation results for (a)~biaxially compressed sheets with periodic boundary conditions, (b)~uniaxially compressed sheets with periocic boundary conditions, and (c)~uniaxially compressed sheets with clamped-free boundary conditions. Snapshots on the left show typical sheet configurations pre- and post-buckling. For clarity, the height profiles of sheets ($z$ coordinates) are also indicated by a heat map, where yellow indicates the largest heights and dark blue
indicates the lowest heights. Plots in the middle and right columns show characteristic stress-strain curves at low temperature ($\ell_0/\ell_\text{th} =0.08$) and at high temperature ($\ell_0/\ell_\text{th}=2,500$), respectively, where stresses are averaged over 30 simulation runs. Red dashed lines show linear fits to the first 600 (pre-buckling) and to the last 900 points (post-buckling) out of total 3,000 points.
For periodic boundary conditions we plot relative strains $\epsilon = (\langle\ell_x(\sigma=0) \rangle -\ell_{x})/\ell_{0x}$, where $\ell_{0x}$ and $\ell_{x}$ are the undeformed and the deformed lengths of the sheet, respectively, and $\langle\ell_x(\sigma=0) \rangle$ corresponds to the equilibrium length of the sheet at zero load. For clamped boundary conditions we plot absolute strains $\epsilon = (\ell_{0x} -\ell_{x})/\ell_{0x}$.
Stresses $\sigma$ are normalized with the zero temperature critical buckling load $\sigma_0^b = 4 \pi \kappa_0 \ell_0^{-2}$. Strains $\epsilon$ are normalized with the zero temperature critical buckling strains $\epsilon_0^b$, which are $\epsilon_0^b=\sigma_0^b/Y_0$ for the uniaxial compression and $\epsilon_0^b=\sigma_0^b/(2 B_0)$ for the biaxial compression, where $B_0=Y_0/[2 (1-\nu_0)]$ is the 2D bulk modulus and $\nu_0=1/3$ is the Poisson's ratio.}
\label{fig3}
\end{figure*}

Figure~\ref{fig3} shows typical stress-strain curves at low temperature ($\ell_0/\ell_\text{th} =0.08$) and at high temperature ($\ell_0/\ell_\text{th}=2,500$) for 3 different loading and boundary conditions, where  compressive strains were prescribed and were gradually increased in 3,000 increments, while compressive stresses were calculated using the virial stress equation and averaged over 30 independent simulation runs (see Appendix for details). Note that for the biaxially and uniaxially compressed sheets with periodic boundary conditions we plotted relative strains that are calculated relative to the projected equilibrium sheet length $\langle \ell_x (\sigma=0) \rangle$ at zero load, which was obtained with an initial NPT simulation (see Appendix). The projected equilibrium sheet length $\langle \ell_x (\sigma=0) \rangle$ is smaller than the undeformed sheet length $\ell_{0x}$ due to the out-of-plane fluctuations~\cite{de2012bending,kosmrlj16,amorim16}. For the clamped boundary condition we were unable to perform NPT-like simulations. Thus we plotted absolute strains that are calculated relative to the undeformed sheet length $\ell_{0x}$. Note that the sheet is under tension at zero absolute strain (negative stress values in Fig.~\ref{fig3}c) because thermal fluctuations cause shrinking of the sheet, which has to be then pulled back to the initial length.

At low temperatures ($\ell_0/\ell_\text{th}=0.08$ in Fig.~\ref{fig3}) we recover classical results~\cite{timoshenkoBShells,koiterB}. In the pre-buckling regime, the slopes for the uniaxially and biaxially compressed sheets are equal to $Y_0$ and $2 B_0$, respectively, where $B_0=Y_0/[2 (1-\nu_0)]$ is the 2D bulk modulus and $\nu_0=1/3$ is the Poisson's ratio. In the post-buckling regime, the slope is zero for the uniaxially compressed sheets, but non-zero for the biaxially compressed sheets. This is because at the critical buckling load $\sigma_0^b = 4 \pi \kappa_0 \ell_0^{-2}$ only one mode (${\bf q}_1=(2\pi/\ell_{0x},0)$) becomes unstable  for the uniaxial compression (see Fig.~\ref{fig3}b,c), while two modes (${\bf q}_1=(2\pi/\ell_{0x},0)$ and ${\bf q}_2=(0,2\pi/\ell_{0y})$) become unstable for the biaxial compression (see Fig.~\ref{fig3}a). The linear combination of the two unstable modes produces Gaussian curvature, which forces the sheet to stretch and the resulting slope is reduced to $B_0/2$~\cite{timoshenkoBShells,koiterB}. Note that at low temperatures we observe periodic oscillations in the stress-strain curves (see $\ell_0/\ell_\text{th}=0.08$ in Fig.~\ref{fig3}). This is because the auto-correlation times for the soft long wavelength modes are very long and we were unable to fully equilibrate the sheet.

At high temperatures ($\ell_0/\ell_\text{th}=2,500$ in Fig.~\ref{fig3}) the stress-strain curves exhibit larger fluctuations because the amplitude of thermal fluctuations is increased, but we can still identify two different regimes (red dashed lines in Fig.~\ref{fig3}) at low strains and large strains, which correspond to the pre-buckling and post-buckling regimes. The slopes in the pre-buckling regime for the uniaxially and biaxially compressed sheets correspond to the renormalized Young's modulus and bulk modulus, respectively. Note that the renormalized Young's modulus and bulk modulus are reduced compared to the zero temperature values (see Fig.~\ref{fig1}), which is reflected in the fact that slopes are less than 1 in the normalized stress-strain curves in Fig.~\ref{fig3}. Furthermore, the critical buckling load (intersection of two red dashed lines) is significantly increased compared to the zero temperature value as we predicted in Eq.~(\ref{eq:normalized_buckling_load}).

\begin{figure}[t]
\centering
\includegraphics[width=0.45\textwidth]{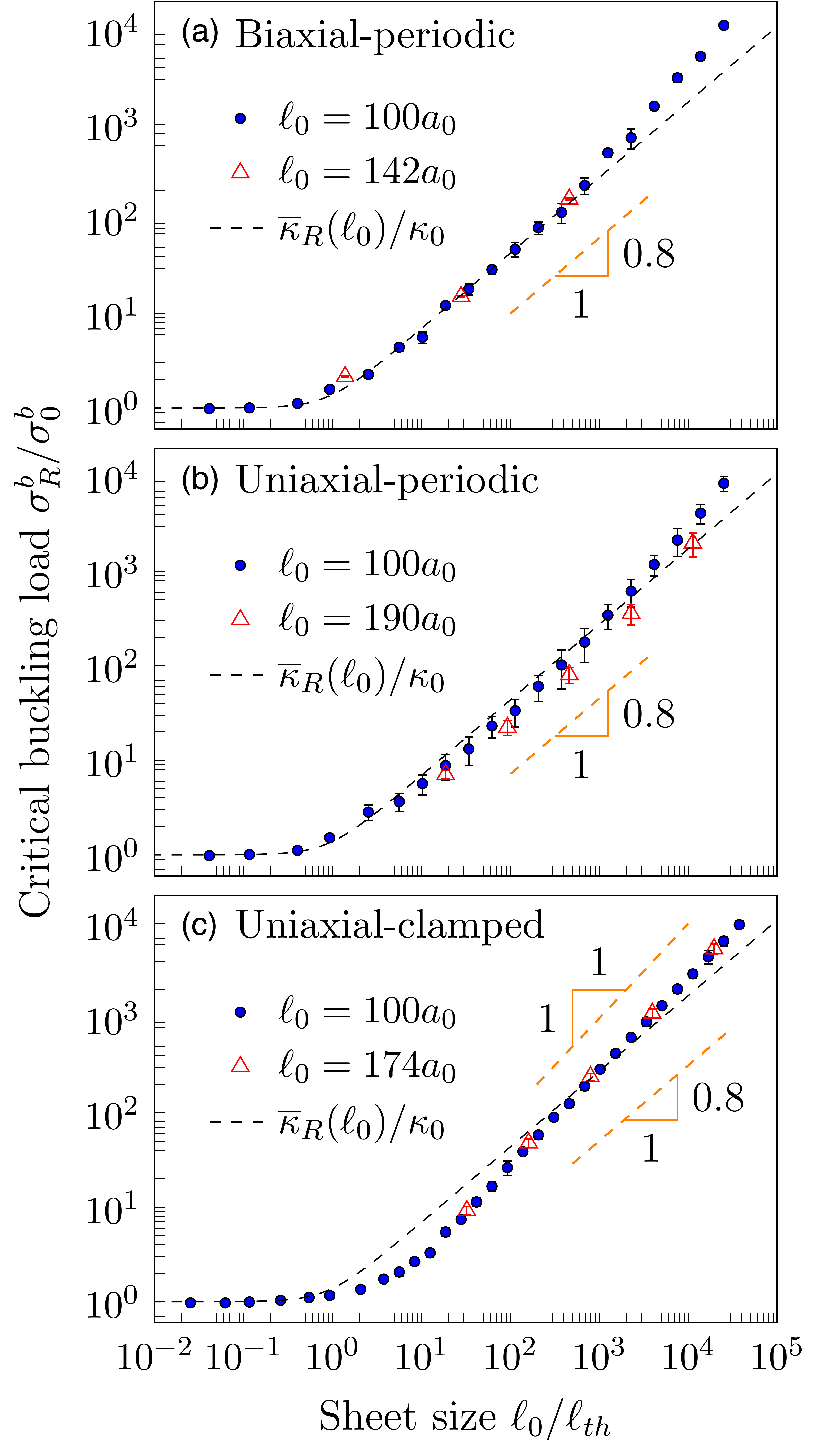}
\caption{Scaling functions for the critical buckling load $\sigma_R^b$ for (a)~biaxially compressed sheets with periodic boundary conditions, (b)~uniaxially compressed sheets with periocic boundary conditions, and (c)~uniaxially compressed sheets with clamped-free boundary conditions. The critical buckling loads $\sigma_R^b$ and confidence intervals (errorbars) were obtained from the stress-strain curves in Fig.~\ref{fig3}. Simulations were done for two different sheet sizes $\ell_0$ (blue dots and red triangles) and thermal length scale $\ell_\text{th}$ defined in Eq.~(\ref{eq:thermal_length}) was tuned by varying the temperature $T$, and the spring constants $k_\text{bend}$ and $k_\text{stretch}$ for the bending and stretching. The critical buckling loads $\sigma_R^b$ were normalized by the classical zero temperature value $\sigma_0^b = 4 \pi \kappa_0 \ell_0^{-2}$ and plotted as a function of the normalized sheet size $\ell_0/\ell_\text{th}$. Dashed black line shows the theoretical prediction from Eq.~(\ref{eq:normalized_buckling_load}).}
\label{fig4}
\end{figure}

Finally, in Figure~\ref{fig4} we  compare the critical buckling load $\sigma_R^b$ obtained from the stress-strain curves (see Fig.~\ref{fig3}) with the theoretical predictions in Eq.~(\ref{eq:normalized_buckling_load}). For the biaxially compressed sheets the critical buckling loads and the confidence intervals were obtained by the cross-sections of two linear lines that correspond to the fits for the pre- and post-buckling regime (see Fig.~\ref{fig3}a). For the uniaxially compressed samples, the value of stress levels off in the post-buckling regime (see Fig.~\ref{fig3}b,c). Thus we estimated the critical buckling load and confidence intervals by doing long simulations at the maximum compressive strain in the post-buckling regime (see Appendix for details).

By varying the size of the sheet $\ell_0$, the ambient temperature $T$, and the spring constants $k_\text{bend}$ and $k_\text{stretch}$ for the bending and stretching, respectively, we are able to tune the ratio of $\ell_0/\ell_\text{th}$ by 7 orders of magnitude ($\ell_0/\ell_\text{th}\in (10^{-2},10^5)$). All the values for critical buckling loads can be collapsed to a single scaling function as predicted by the theory in Eq.~(\ref{eq:normalized_buckling_load}).

For the biaxially and uniaxially compressed sheets with periodic boundary conditions we get a remarkably good agreement with the theoretical predictions, which confirms that thermal fluctuations increase the critical buckling load due to the enhanced renormalized bending rigidity. Recall that we were unable to fully equilibrate simulations at low temperatures (oscillations for $\ell_0/\ell_\text{th}=0.08$ in Fig.~\ref{fig3}), but in this regime the effects of thermal fluctuations are small and the critical buckling loads $\sigma_R^b$ are well approximated by the classical critical buckling load $\sigma_0^b=4 \pi \kappa_0 \ell_0^{-2}$. At very large temperatures ($\ell_0/\ell_\text{th} \gtrsim 10^4$) we observe systematic deviation from the theoretical prediction. Inspection of the sheet configurations revealed that the local bond strains fluctuate by $\pm5-10\%$ and that the local dihedral angles fluctuate by $\pm 20^\circ$. For such large fluctuations the discretized energy in Eq.~(\ref{eq:energy}) starts deviating from the linear elastic sheet that was assumed for theoretical calculation. Furthermore, according to the Lindemann criterion~\cite{lindemann10,born1939thermodynamics}, such large fluctuations would cause the sheet to melt, which was prevented in our simulations, where the connectivity between particles was fixed. Melting of fluctuating sheets is still an unresolved problem and we leave this for future work.

For the uniaxially compressed sheets with clamped boundary condition, we also observe a scaling collapse of the critical buckling load (Fig.~\ref{fig4}c). However, we notice a systematic deviation from the theoretical prediction in Eq.~(\ref{eq:normalized_buckling_load}). In previous studies of thermalized ribbons it was noted that the effect of clamped boundaries decays in the bulk with the scale that is of the order of the width of the ribbon~\cite{bowick17,wan2017thermal,russell2017stiffening}. Thus the effect of clamped boundaries is felt throughout the square sheets, which affects the renormalization of the bending rigidity in the bulk. Nonetheless, in this case we still see that thermal fluctuations can significantly increase the critical buckling load for large values of $\ell_0/\ell_\text{th}$.


The results presented above are universal and they hold for any free-standing elastic sheet, where the continuum theory applies, i.e. when the sheet is much larger than the microscopic cutoff, e.g. the interatomic spacing in 2D crystalline sheets. For sheets that are much smaller than the thermal length scale $\ell_\text{th}\sim{\kappa_0/\sqrt{k_B T Y_0}}$ thermal fluctuations are negligible and classical mechanics applies. This is the case for all macroscopic sheets and plates because the thermal length scale rapidly increases with the sheet thickness $t$ and scales as $\ell_\text{th}\sim t^{5/2} E^{1/2} (k_B T)^{-1/2}$, where $E$ is the 3D Young's modulus ($\kappa_0 \sim E t^3$, $Y_0 \sim E t$). As the sheet thickness is reduced to the order of nanometers, as is the case for clay plates and red blood cells, the thermal length scale $\ell_\text{th}$ becomes comparable to the size $\ell_0$ of the sheet. In this regime thermal fluctuations become relevant and the renormalized bending rigidity and the critical buckling load are mildly increased (Fig.~\ref{fig4}). Note that in red blood cells the molecular activity produces non-equilibrium fluctuations, which dominate over the thermal fluctuations on large length scales as was deduced from the breakdown of the fluctuation-dissipation theorem~\cite{turlier16}. In the future it would thus be worth exploring how such dynamic non-equilibrium fluctuations affect the mechanical properties of slender structures.

As the sheet thickness is reduced to the atomistic scale, such as for graphene, boron nitride, transition metal dichalcogenide, and other 2D materials, the thermal length scale becomes of the order of nanometers at room temperature, which is much smaller than the characteristic size of these sheets. In this regime thermal fluctuations significantly increase the critical buckling load ($\ell_0/\ell_\text{th} \gg 1$ in Fig.~\ref{fig4}), which becomes temperature dependent and scales as $\sigma_R^b \sim \kappa_0 \ell_0^{-2+\eta} \ell_\textrm{th}^{-\eta} \sim \kappa_0^{1-\eta} \ell_0^{-2+\eta} (k_B T Y_0)^{\eta/2}$. In this letter we focused on pristine elastic sheets, but defects are often unavoidable in 2D materials and they produce static ripples. It was previously demonstrated that quenched defects also enhance the bending rigidity and they can dominate over the thermal fluctuations when the amplitudes of static ripples is larger than the amplitude of height fluctuations due to temperature
~\cite{nelson91,radzihovsky91,morse92,ledoussal93,kosmrlj13,kosmrlj14,gornyi15,le2018anomalous}. Since the critical buckling load studied in this work scales with the effective bending rigidity, we expect that defects and static ripples will also increase the critical buckling load. Moreover, the static ripples and other mechanical deformations in 2D materials can affect their electronic transport~\cite{mariani_prl08,castro_prl10,mariani_prb10,guinea_prb08,guinea_SSC09,castroneto_RevModPhys09,amorim16}. The coupling between elastic deformations and the electronic degrees of freedom could also affect the effective mechanical behavior of suspended membranes~\cite{gazit_prb09,PhysRevE.80.041117,sanjose_prl11,guinea_prb14,G14}. In particular, they could produce spontaneous buckling~\cite{bonilla2016critical,ruiz2016stm,ruiz2017bifurcation} and stable ripples~\cite{cea2019large,cea2019numerical,ruiz2015ripples}.  

We hope this letter will stimulate further experimental, numerical, and theoretical investigations of the stability and
mechanical properties of thermalized sheets as well as extensions to more complicated geometries found in microscopic kirigami and origami structures.

This work was supported by NSF through the Career Award DMR-1752100 (A.K.), the CMMI Grant No. 1825921 (M.J.A.Q.), and the DMR Grant No. 1506625 (M.R.-G.). M.R.-G. also acknowledges support from the Simons Foundation via awards 327939 and 454945.

\appendix

\section{Molecular Dynamics simulations}
\label{app:simulations}

All Molecular Dynamics simulations were performed using the LAMMPS package~\cite{LAMMPS}. We chose the lattice constant $a_0$, the particle mass $m$, and $k_BT$ as the fundamental units for length, mass, and energy, respectively. The units of time and stress were $\tau=a_0\sqrt{m/k_BT}$ and $\sigma=k_BT/a_0^2$, respectively. The velocity Verlet algorithm was used for the integration of equations of motion with a timestep of $\Delta t=0.005\tau$ and Nos\'e-Hoover thermostat and barostat~\cite{tuckerman2010statistical} were used to control the ambient temperature and pressure, where needed. For all simulations we fix the box height in the $z$-direction, where periodic boundary conditions were used. In all simulations stress components were computed using the virial stress equation~\cite{tuckerman2010statistical}.

For simulations with periodic boundary conditions we initially equilibrated the sheet by doing $2\times10^6$ timesteps of the NPT simulation at zero pressure such that $\langle \sigma_{xx}\rangle = \langle \sigma_{yy}\rangle=0$. After that we gradually strain the system in 3,000 increments with 5,000 timesteps between each increment to equilibrate the sheet. Our sensitivity analysis showed that doubling the equilibration time interval after each strain increment did not affect the results. For the biaxial compression we strain the system by prescribing the reduced box size in both $x$- and $y$-directions. For the uniaxial compression we strained the system by prescribing the reduced box size in the $x$-direction, while the box size in the $y$-direction was allowed to fluctuate such that $\langle \sigma_{yy}\rangle=0$.  

For simulations with clamped boundary conditions we fixed two rows of particles along each of the two opposite edges. These particles were clamped in the $x$- and $z$-directions, but they were free to move in the $y$-direction. The two other edges were free to move and they were independent, i.e. they were not connected with periodic boundary condition. For these sets of simulations we gradually strained the system by bringing the two clamped edges closer together in 3,000 increments with 5,000 timesteps between each increment to equilibrate the compressed sheet.

The post-buckling stress for the uniaxially compressed sheets with periodic and clamped boundary conditions were calculated by performing $2 \times 10^7$ timesteps at the maximum compressive strain and the average compressive stress was estimated using the virial stress equation.

\bibliography{library.bib}

\end{document}